# *Coupling and Guided Propagation along Parallel Chains of Plasmonic Nanoparticles*

Andrea Alù[1,*], Pavel A. Belov[2,3], and Nader Engheta[4]

[1]Department of Electrical and Computer Engineering, The University of Texas at Austin, Austin, TX 78712, USA

E-mail: alu@mail.utexas.edu

[2]Department of Electronic Engineering, Queen Mary University of London, Mile End Road, London, E1 4NS, United Kingdom

E-mail: pavel.belov@elec.qmul.ac.uk

[3]Department of Photonics and Optoinformatics, St. Petersburg State University of Fine Mechanics and Optics

Kronverksky Pr. 49, 197101, St. Petersburg, Russia

E-mail: belov@phoi.ifmo.ru

[4]Department of Electrical and Systems Engineering, University of Pennsylvania

200 South 33rd Street – ESE 203 Moore

Philadelphia, PA 19104, U.S.A., tel. +1.215.898.9777

E-mail: engheta@ee.upenn.edu

---

[*] To whom correspondence should be addressed: Email: alu@mail.utexas.edu

*Abstract*

Here, extending our previous work on this topic, we derive a dynamic closed-form dispersion relation for a rigorous analysis of guided wave propagation along coupled parallel linear arrays of plasmonic nanoparticles, operating as optical "two-line" waveguides. Compared to linear arrays of nanoparticles, our results suggest that these waveguides may support longer propagation lengths and more confined beams, operating analogously to transmission-line segments at lower frequencies. Our formulation fully takes into account the whole dynamic interaction among the infinite number of nanoparticles composing the parallel arrays, considering also realistic presence of losses and the frequency dispersion of the involved plasmonic materials, providing further physical insights into the guidance properties that characterize this geometry.

**1. Introduction**

Linear chains of plasmonic (silver or gold) nanoparticles have been suggested as optical waveguides in several recent papers [1]-[11]. Owing to design flexibility and relatively easy construction within current nanotechnology, the realization of such ultracompact waveguides has been thoroughly studied and analyzed in the past few years. However, the recent experimental realizations of such devices at the nanoscale have revealed challenges due to severe sensitivity to material absorption and to inherent disorder. The guided beam cannot usually travel longer than few nanoparticles before its amplitude is lost in the noise. This is mainly due to the fact that linear arrays of small nanoparticles have the property to concentrate the optical beam in a narrow region of space, in large part filled by lossy metal. If this is indeed appealing in terms of power concentration, it has the clear disadvantage of strong sensitivity to material and radiation losses.

As we have underlined in [12], a naked conducting wire at low frequencies has analogous limitations: although metals are much more conductive and less lossy in radio frequencies, connecting two points in a regular circuit with a single wire would still produce unwanted spurious radiation and sensitivity to metal absorption. This problem, which is much amplified at optical frequencies due to the poorer conductivity and higher loss of metals in the visible, is simply approached at low frequencies by closely pairing two parallel wires (or, which is the same, placing a ground plane underneath the conducting trace), forming the well known concept of a transmission-line that provides a return path for the conduction current. Analogously, applying the nanocircuit concepts [13]-[14], we have recently put forward ideas to realize optical nanotransmission-line waveguides in different geometries [15]-[16], which have been proven to be more robust to material and radiation losses and may provide wider bandwidth of operation. In particular, as we introduced in [12], one such idea consists in pairing together two parallel arrays of plasmonic nanoparticles, suggesting that the coupling among the guided modes may improve the guidance performance. In [12] we have shown that this is indeed the case: operating with the antisymmetric longitudinal mode, such parallel chains indeed may confine the beam in the background region between the chains, leading to confined propagation that is combined with robustness to material absorption and radiation losses. In particular, we have shown that operating with these modes near the light-line would, in many senses, lead to operation close to a regular transmission-line at low frequencies, but available in the visible regime.

Here, we extend our work in this area by deriving a closed-form full-wave dynamic solution for the dispersion of the eigenmodes supported by such parallel chains, fully taking into account the coupling among the infinite number of particles composing the two-chain array, even in the presence of material absorption, radiation losses and frequency dispersion. The results confirm the validity of this analogy, and they provide further insights into the operation and spectrum of modes guided by these paired arrays of nanoparticles. Applications for low-loss optical communications and sub-wavelength imaging devices are envisioned.

## 2. Dispersion Relations for Guided Propagation

Consider the geometry of Fig. 1, i.e., two identical linear arrays of plasmonic nanoparticles with radius $a$, period $d > 2a$ and interchain distance $l > d$. This geometry has been preliminarily analyzed in [12] for its longitudinally polarized guided modes, where it was shown that the coupling between the chains, limited in that analysis to its dominant contribution coming from the averaged current density on the chain axes, would generate the splitting of the regular longitudinal mode into two coexisting longitudinal modes, respectively, with symmetric and antisymmetric field distributions. The antisymmetric mode is the one corresponding to transmission-line operation [12], as outlined in the introduction, for which two antiparallel displacement current flows are supported by the parallel chains. A similar modal propagation has been analyzed in [9] for a related distinct geometry, consisting of longitudinal dipoles placed over a perfectly conducting plane. Also our analysis of quadrupolar chains [16] may, in the limit of $l \to 0$, have some analogies with this antisymmetric operation. In the following, we rigorously approach the general problem of modal dispersion along the parallel chains of Fig. 1, extending our general analysis in [10] that was valid for one isolated chain. Our formulation may fully take into account the whole coupling among the infinite nanoparticles composing the pair of arrays and the possible presence of material absorption, radiation losses and frequency dispersion.

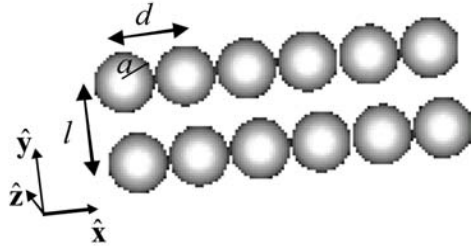

Figure 1 – (Color online). Geometry of the problem: a pair of linear arrays of plasmonic nanoparticles as an optical two-line waveguide.

As we did in [12], here we model each nanoparticle as a polarizable dipole with polarizability $\alpha$, an assumption that is valid as long as $a \ll \lambda_b$, with $\lambda_b$ being the wavelength of operation in the background material. For simplicity, we assume a scalar polarizability, implying that the particles are isotropic (nanospheres, easy to realize as colloidal metal particles), or for more general shapes focusing on one specific field polarization. In the following, we also assume an $e^{-i\omega t}$ time convention.

For a single isolated chain [10], the spectrum of supported eigenmodes may be split into longitudinal and transverse polarization with respect to the chain axis $\hat{\mathbf{x}}$. In particular, for $e^{i\beta x}$ propagation, the corresponding guided wave number $\beta$ satisfies the following closed-form dispersion relations, respectively, for longitudinal and transverse modes:

$$L = 3\bar{d}^{-3}\left[f_3(\bar{\beta},\bar{d}) - i\bar{d}\, f_2(\bar{\beta},\bar{d})\right] - \bar{\alpha}^{-1} = 0$$
$$T = -\frac{3}{2}\bar{d}^{-3}\left[f_3(\bar{\beta},\bar{d}) - i\bar{d}\, f_2(\bar{\beta},\bar{d}) - \bar{d}^2 f_1(\bar{\beta},\bar{d})\right] - \bar{\alpha}^{-1} = 0 \quad (1)$$

where $f_N(\bar{\beta},\bar{d}) = Li_N\left(e^{i(\bar{\beta}+1)\bar{d}}\right) + Li_N\left(e^{-i(\bar{\beta}-1)\bar{d}}\right)$, $Li_N(z)$ is the polylogarithm function of order $N$ [17] and all the quantities have been normalized, consistent with [10], as $\bar{d} = k_b d$, $\bar{\beta} = \beta/k_b$, $\bar{\alpha} = k_b^3 \alpha/(6\pi\varepsilon_b)$, with $k_b = 2\pi/\lambda_b$ being the background wave number and $\varepsilon_b$ the corresponding permittivity.

These equations, as shown in details in [10], take fully into account the dynamic coupling among the infinite number of particles composing the linear chain. They are real-valued for lossless particles (for which $\text{Im}\left[\bar{\alpha}^{-1}\right] = -1$ [10]), supporting guided modes with $\bar{\beta} > 1$, but they are also fully valid in the complex domain when realistic material losses are considered, allowing to evaluate the realistic damping factors associated with material absorption and radiation losses. They can be applied also in the leaky-wave modal regime, for which $\text{Re}\left[\bar{\beta}\right] < 1$ and the chain radiates as an antenna in the background region [18].

When $l$ is finite in Fig. 1, the coupling between the two chains implies a modification of their guidance properties, which may be taken into account by considering the polarization fields induced by the electric field from each chain on the other. The fields radiated by each chain may be expanded into cylindrical waves, allowing us to write the general closed-form expressions for the coupling coefficients between the two chains.

Without loss of generality, we can assume that the particles composing the first chain, located at $y = 0$, are polarized by an eigenmodal wave with dipole moments $\mathbf{p}_1 e^{i\beta md}$, where $m \in \Im$ is the integer index for each nanoparticle of the chain. The equivalent current distribution on the $x$ axis may be written as:

$$\mathbf{J}(x) = -i\omega \mathbf{p}_1 \sum_{m=-\infty}^{\infty} e^{i\beta md} \delta(x - md), \qquad (2)$$

where $\delta(.)$ is the Dirac delta function. The fields radiated by such current distribution may be expanded into cylindrical waves and may be used to evaluate the coupling coefficients with between one chain and the other, with dipole moments $\mathbf{p}_2 e^{i\beta md}$ located at $y = l$, yielding:

$$\begin{aligned}
C_{xx} &= -\frac{3}{\bar{d}} \sum_{m=-\infty}^{\infty} \bar{b}_m^2 K_0\left[\bar{b}_m \bar{l}\right] \\
C_{xy} &= C_{yx} = -\frac{3i}{\bar{d}} \sum_{m=-\infty}^{\infty} \sqrt{\bar{b}_m^2 + 1}\, \bar{b}_m K_1\left[\bar{b}_m \bar{l}\right] \\
C_{yy} &= \frac{3}{\bar{l}\bar{d}} \sum_{m=-\infty}^{\infty} \left(\bar{b}_m^2 + 1\right) \bar{l} K_0\left[\bar{b}_m \bar{l}\right] + \bar{b}_m K_1\left[\bar{b}_m \bar{l}\right] \\
C_{zz} &= \frac{3}{2\bar{d}} \sum_{m=-\infty}^{\infty} \left(\bar{b}_m^2 + 2\right) K_0\left[\bar{b}_m \bar{l}\right] - \bar{b}_m^2 K_2\left[\bar{b}_m \bar{l}\right]
\end{aligned} \qquad (3)$$

where $\bar{b}_m = \sqrt{\left(\bar{\beta} + \dfrac{2\pi m}{\bar{d}}\right)^2 - 1}$ and $K_m[.]$ are the modified Bessel functions of order $m$. The generic coupling coefficient $C_{ij}$ expresses the polarization along $j$ on one chain induced by the $i$-polarized dipoles on the other chain. The summations in (3) have very fast convergence, and the dominant term ($m=0$) is usually sufficient to take into account the dominant contribution to the coupling, an approximation that is consistent with the approach we used in [12]. The numerical results reported in the following sections have been obtained by considering the first ten terms in the summations (3), even though full convergence has been usually achieved after the first one or two terms. The other coupling coefficients not explicitly given in (3) are null, implying that longitudinal modes (directed along $x$) and the transverse modes polarized along $y$ are coupled together through $C_{xy}$, whereas transverse modes polarized along $z$ are not coupled with the orthogonal polarizations.

The final closed-form dispersion relation for the eigenmodes supported by the parallel chains may be written as:

$$\det\begin{pmatrix} L & 0 & C_{xx} & C_{xy} \\ 0 & T & C_{xy} & C_{yy} \\ C_{xx} & -C_{xy} & L & 0 \\ -C_{xy} & C_{yy} & 0 & T \end{pmatrix} \det\begin{pmatrix} T & C_{zz} \\ C_{zz} & T \end{pmatrix} = 0, \qquad (4)$$

or, in a more compact form:

$$\left[(L \pm C_{xx})(T \mp C_{yy}) + C_{xy}^{\;2}\right](T \pm C_{zz}) = 0. \qquad (5)$$

The left-hand side in Eqs. (4)-(5) consists of the product of two terms: the first determines the dispersion of the coupled modes polarized in the $xy$ plane (among which the quasi-longitudinal antisymmetric modes that have been considered in [12]), whereas the second determines the purely-transverse modes polarized along $z$. It is noticed that this dispersion equation is completely general and it fully takes into account the whole dynamic interaction among the infinite particles composing the two parallel chains. Since the coupling coefficients (3) tend rapidly to zero for increased $l$, it is noticed that Eq. (5) represents the perturbation of the original transverse and longitudinal modes supported by the two linear chains independently given by $L = 0$ and $T = 0$ respectively [10], produced by the coupling coefficients $C$. In particular, it is seen that each of the three orthogonal polarizations (along $x$, $y$, $z$) splits into two branches due to the coupling between the chains, one with symmetric and the other with antisymmetric properties, leading to six modal branches of guided modes, some of which supported at the same frequency. In particular, the modes in the $xy$ plane are mixed together (i.e., the parallel chains do not support purely longitudinal or purely $y-$polarized modes).

In the limit of lossless particles, since $L$ and $T$ are real for any $\bar{\beta} \geq 1$ [10], by inspecting Eq. (5) we notice that the parallel chains still support lossless guided propagation for any $1 \leq \bar{\beta} \leq \pi/\bar{d}$. In the following, we analyze in details the modal properties of this setup in its different regimes of operation.

## 3. Guided Modes of Parallel Chains of Silver Nanospheres

In this section we consider the different regimes of guided propagation supported by the parallel chains of Fig. 1, considering realistic optical materials composing the plasmonic nanoparticles. In the case of a chain of homogeneous spherical particles of radius $a$ and permittivity $\varepsilon = \varepsilon_r + i\varepsilon_i$, their normalized polarizability satisfies the following relations [10]:

$$\operatorname{Re}\left[\bar{\alpha}^{-1}\right] = \frac{3}{2}(k_b a)^{-3}\frac{\varepsilon_r + 2\varepsilon_b}{\varepsilon_r - \varepsilon_b}$$
$$\operatorname{Im}\left[\bar{\alpha}^{-1}\right] = -1 - \frac{9\varepsilon_i}{2}\frac{\varepsilon_b (k_b a)^{-3}}{(\varepsilon_r - \varepsilon_b)^2 + \varepsilon_i^2}. \qquad (6)$$

Since the guided modes are perturbations of the longitudinal and transverse modes supported by the isolated chains, it is of no need to analyze here again how variations in the chain geometry, i.e., in $a$, $d$ and/or the involved materials, may affect the guidance of the parallel chains, since in [10] we have already studied in great details how these changes affect the guidance of isolated chains In the following, therefore, we focus on one specific realistic design of the chains and we employ the exact formulation developed in the previous section to characterize the modal properties of two of such parallel arrays coupled together. In particular, the geometry of interest is formed by colloidal silver nanospheres embedded in a glass background ($\varepsilon_b = 2.38\varepsilon_0$). We use experimental data available in the literature to model the silver permittivity at optical frequencies [21] and we assume $a = 10\,nm$ and $d = 21\,nm$ for the two chains.

*a) Quasi-longitudinal propagation (forward modes)*

As we have shown in [10], an isolated linear chain of plasmonic nanoparticles supports forward-wave longitudinal guided modes ($x-$ polarized), satisfying the dispersion relation $L = 0$, over the frequency regime for which:

$$6\left[Cl_3(\bar{d}+\pi)+\bar{d}\,Cl_2(\bar{d}+\pi)\right] < \bar{d}^3 \operatorname{Re}\left[\bar{\alpha}^{-1}\right] < 3\left[\xi(3)+Cl_3(2\bar{d})+\bar{d}\,Cl_2(2\bar{d})\right],$$

(7)

where $Cl_N(\theta)$ are Clausen's functions [17]. For the case at hand (silver nanoparticles, $a = 10\,nm$ and $d = 21\,nm$), such modal regime is supported over a relatively wide range of frequencies between $550\,THz$ and $850\,THz$, as reported in Fig. 2 (thin solid black line). In particular, in the figure we plot the real and imaginary parts of the normalized $\bar{\beta}$ and the propagation length, i.e., the distance traveled by the guided mode before its amplitude is $e^{-1}$ of the original value, which is equal to $\operatorname{Im}[\beta]^{-1}$. The shadowed regions at the sides of the plots delimit the leaky-wave region (left-side, lighter blue shadowed region), for which $\operatorname{Re}[\bar{\beta}] < 1$ and the mode radiates in the background region, and the stop-band region (right side, darker shadow, brown), for which when lossless particles are considered $\operatorname{Re}[\bar{\beta}] = \pi / \bar{d}$ and the mode is evanescent in nature. In between these two regions, as defined by Eq. (7), the modes are guided and $\operatorname{Im}[\beta]$, i.e., the damping factor, is only associated with material losses, since in the limit of lossless particles the mode would not radiate and $\operatorname{Im}[\beta] = 0$ [10]. In the leaky-wave region the damping is larger, due to radiation losses [18], whereas in the stop-band the mode does not propagate and it is reflected back by the chain due to Bragg reflection. Near the light line ($\text{Re}[\bar{\beta}] \simeq 1$) the mode is poorly guided by an isolated chain, but its propagation length may reach relatively large values, around $1 \mu m$.

Consider now the case for which $l$ is finite, i.e., the two parallel chains are coupled together. In this case, the longitudinal modes are coupled with each other, also polarizing the chains with a small transverse polarization along $y$, consistent with the value of $C_{xy}$. The longitudinal mode dispersion splits into two quasi-longitudinal branches, one with symmetric and the other with antisymmetric properties with respect to $x-$ polarization. The two modes satisfy, respectively, the following dispersion relations, consistent with Eq.(5):

$$\begin{aligned} sym: & \quad (L+C_{xx})(T-C_{yy})+C_{xy}^{2}=0 \\ antisym: & \quad (L-C_{xx})(T+C_{yy})+C_{xy}^{2}=0 \end{aligned}, \quad (8)$$

providing the following constraints on the polarization eigenvectors for the two chains:

$$\begin{aligned} sym: & \quad \begin{cases} \mathbf{p}_1 \cdot \hat{\mathbf{x}} = \mathbf{p}_2 \cdot \hat{\mathbf{x}} \\ \mathbf{p}_1 \cdot \hat{\mathbf{y}} = -\mathbf{p}_2 \cdot \hat{\mathbf{y}} \end{cases} \\ antisym: & \quad \begin{cases} \mathbf{p}_1 \cdot \hat{\mathbf{x}} = -\mathbf{p}_2 \cdot \hat{\mathbf{x}} \\ \mathbf{p}_1 \cdot \hat{\mathbf{y}} = \mathbf{p}_2 \cdot \hat{\mathbf{y}} \end{cases} \end{aligned}. \quad (9)$$

Figure 2 reports as a first example the dispersion of symmetric and antisymmetric modes for $l = 50 nm$. It is noticed that the small coupling between the chains slightly perturbs the dispersion of the modes, causing the antisymmetric mode (blue dashed line, with polarization currents oppositely flowing along the chains) to have slightly larger real and imaginary parts of $\bar{\beta}$ with respect to the unperturbed longitudinal mode supported by an isolated chain (light solid line). Conversely, the symmetric mode (thick red solid line) supports slightly lower values of $\text{Re}[\bar{\beta}]$. The perturbation is stronger near the light line and in the leaky-wave region, since the mode is less confined around each chain in this regime. The symmetric operation allows an increase of the propagation length of up to $1.5\,\mu m$, since the coupling between the parallel chains with polarization currents flowing in the same direction can boost up the mode. On the other hand, the antisymmetric operation has slightly lower propagation lengths, but this is accompanied with the important advantage of much stronger field confinement, as we highlight in the following. The derivative $\partial\text{Re}[\bar{\beta}]/\partial\omega > 0$ ensures that the modes supported in this regime are all forward-wave, and this is also confirmed by the condition $\text{Im}[\bar{\beta}] > 0$, which ensures that phase and group velocity are parallel with each other for both modes.

As an aside, it should be noted that in the leaky-wave region (blue lighter shadow in the left) the forward-wave modes are improper in nature [22], implying that the dominant cylindrical wave radiated by the chain grows with the distance from the chain instead of decaying. This implies that for a correct evaluation of the modal properties and the field distribution generated in this forward-leaky mode regime, the formulas of Eq. (3) for the index $m = 0$ need to be corrected to the corresponding Hankel functions of second order.

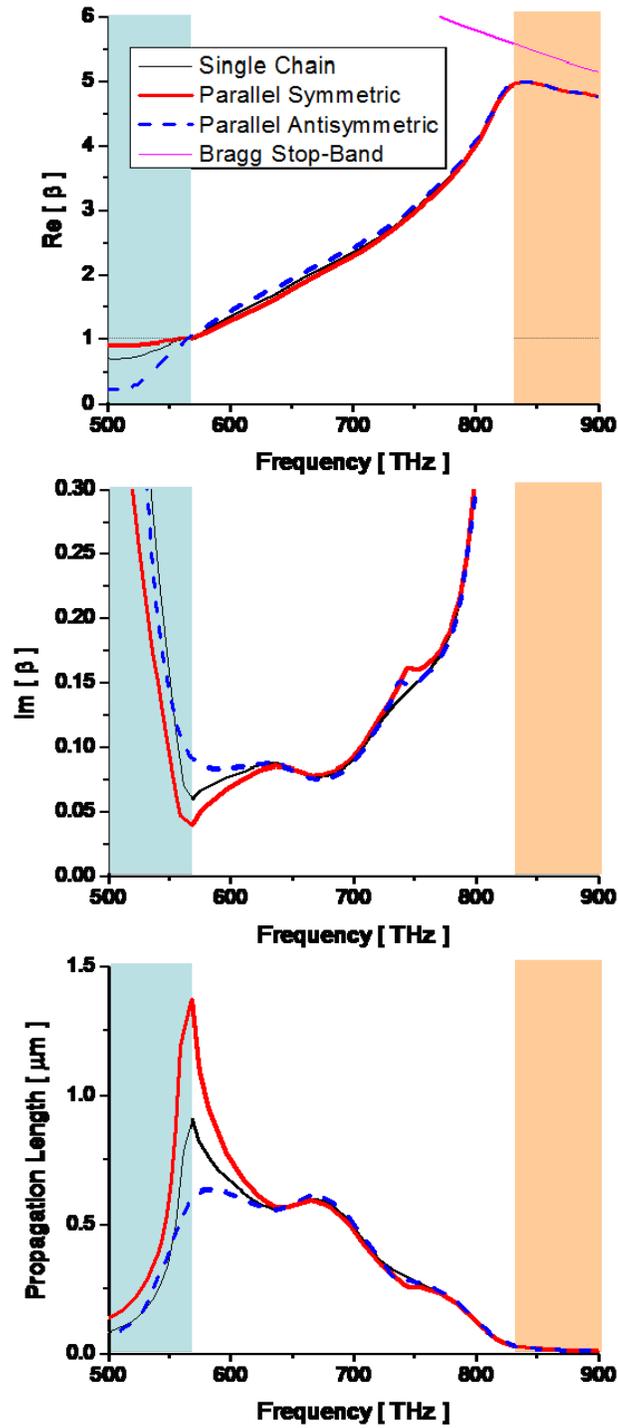

Figure 2 – (Color online). Modal dispersion for the modes supported by two parallel chains with interchain distance $l = 50\,nm$. The dispersions are compared to that of an isolated chain.

Figure 3 reports analogous results for closer chains, with $l = 30\,nm$. It is seen that the perturbation from the isolated chain is now stronger and the coupling between the modes generates some isolated resonant regions of stronger absorption. Still, near the light line propagation lengths are relatively large.

Figure 4 reports the orthogonal magnetic field distribution (snapshot in time) on the $xy$ plane for the modes supported by the chains of Fig. 3 ($l = 30\,nm$) at the frequency $f = 585\,THz$, near the light line. The figure highlights how the modal distribution is quite different in the three scenarios, even if the guided wave numbers are similar. Fig. 4a corresponds to antisymmetric propagation, for which the two chains support the eigenvector polarizations $\mathbf{p}_1 = \hat{\mathbf{x}} + (0.14i - 0.008)\hat{\mathbf{y}}$, $\mathbf{p}_2 = -\hat{\mathbf{x}} + (0.14i - 0.008)\hat{\mathbf{y}}$, consistent with Eq. (9). The corresponding normalized wave number at this frequency is $\bar{\beta}_{asym} = 1.38 + i\,0.1$. It can be seen how the magnetic field is very much confined in the tiny background region delimited by the two chains, similar to the field propagation in a regular transmission-line at low frequencies. Also the electric field is mainly transverse in the region between the chains, supporting the transverse electromagnetic configuration, again typical of a transmission-line mode. This regime of operation, whose interesting properties we have already highlighted in details in [12], may lead to ultra-confined low-loss optical guidance in terms of optical nanotransmission-lines.

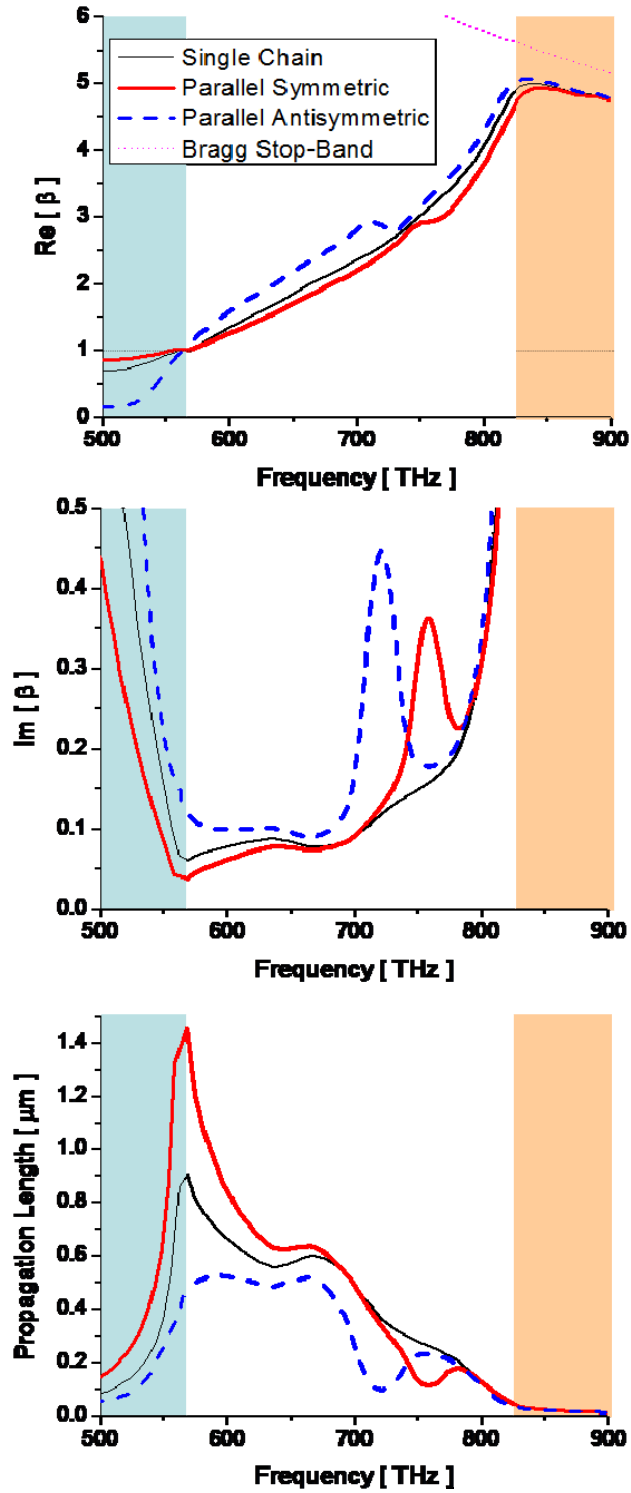

Figure 3 – (Color online). Similar as in Fig. 2, but for $l = 30\,nm$.

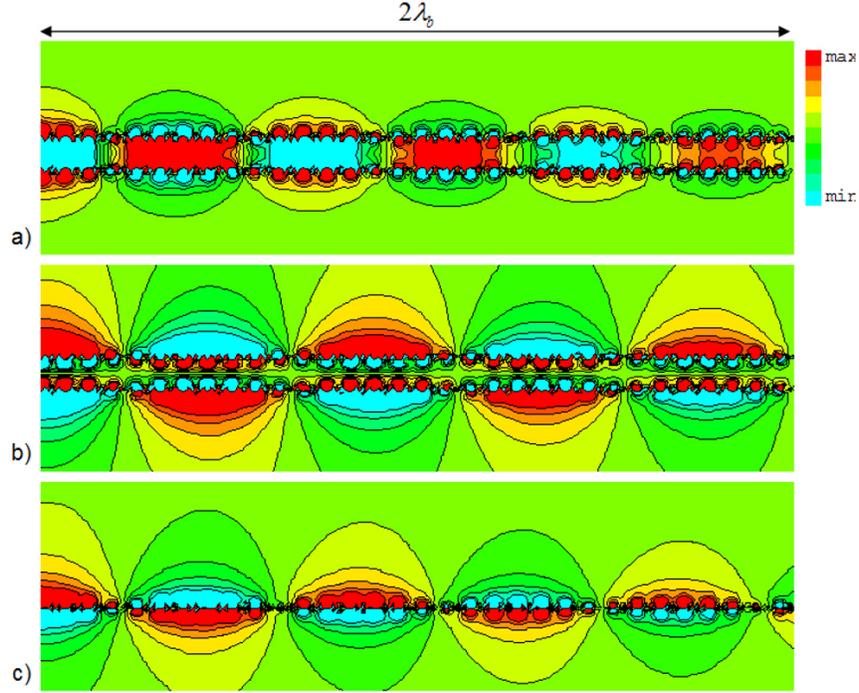

Figure 4 – (Color online). Magnetic field distribution (snapshot in time) for the chains of Fig. 3 at frequency $f = 585 THz$. (a) antisymmetric mode, (b) symmetric mode, (c) isolated chain. All the plots are drawn with the same color scale bar (normalized to the modal amplitude at the left of the figure). The total length of the simulated region is $2\lambda_b$.

Fig. 4b, on the other hand, refers to the symmetric mode for the same parallel chains. In this case, $\mathbf{p}_1 = \hat{\mathbf{x}} + (0.08i - 0.003)\hat{\mathbf{y}}$, $\mathbf{p}_2 = \hat{\mathbf{x}} - (0.08i - 0.003)\hat{\mathbf{y}}$ and $\bar{\beta}_{sym} = 1.13 + i0.053$. The currents flowing along the chains are now parallel with each other, producing fields very much spread all around the outside background region and weak field concentration in between them. This operation is equivalent to two parallel current flows, leading to small fields in between them. Analogous guidance is offered by a single linear chain, reported in Fig. 4c (for comparison, in this third example the chain is positioned at the same location as the lower

chain in the other two panels). In this case the mode is purely longitudinal and $\bar{\beta}_{single} = 1.18 + i\,0.072$, implying weak guidance.

Comparing the three field plots (notice that for fair comparison they have been calculated with the same color scale and under the same initial amplitude excitation), it becomes evident that the antisymmetric longitudinal operation allows a much stronger confinement of the field, with comparable propagation length. Increasing the distance between the chains, as in the examples of Fig. 2, would achieve similar confinement in the region between the chains with reduced attenuation.

These properties are not only limited to the modes operating near the light line, but they are also valid for higher frequencies and more confined modes. For instance, in Fig. 5 we have reported the magnetic field plots for the same chains, operating this time at $f = 680 THz$. At these frequencies, as seen in Fig. 3, the three cases have similar levels of absorption and more confined slow-wave modes. The antisymmetric excitation is characterized in this case by $\mathbf{p}_1 = \hat{\mathbf{x}} + (0.37i - 0.038)\hat{\mathbf{y}}$, $\mathbf{p}_2 = -\hat{\mathbf{x}} + (0.37i - 0.038)\hat{\mathbf{y}}$ and $\bar{\beta}_{asym} = 2.42 + i\,0.091$. Its field distribution (Fig. 5a) still shows strong confinement between the two chains, where a "quasi-uniform" magnetic field may propagate as if guided by a transmission-line. The wave is slower than in the case of Fig. 4, due to increased $\mathrm{Re}[\bar{\beta}]$, but the level of absorption is still quite good and the mode can propagate for over two wavelengths with no strong attenuation. The symmetric operation, for which $\mathbf{p}_1 = \hat{\mathbf{x}} - (0.158i - 0.082)\hat{\mathbf{y}}$, $\mathbf{p}_2 = \hat{\mathbf{x}} + (0.158i - 0.082)\hat{\mathbf{y}}$ and $\bar{\beta}_{sym} = 1.9 + i\,0.074$, once again provides worse field confinement, as expected. In this case (Fig. 5b) the field is spread around the chains and is very weak in the region between the two chains. Similar spreading is noticeable in the single isolated chain configuration of Fig. 5c, with $\bar{\beta}_{single} = 2.06 + i\,0.077$. We note that the field spreading in the region around the chains would also be more sensitive to radiation losses produced by disorder and technological imperfections. We predict, therefore, that the antisymmetric transmission-line operation of the parallel chains may produce more robust optical guidance confined in the region between the chains.

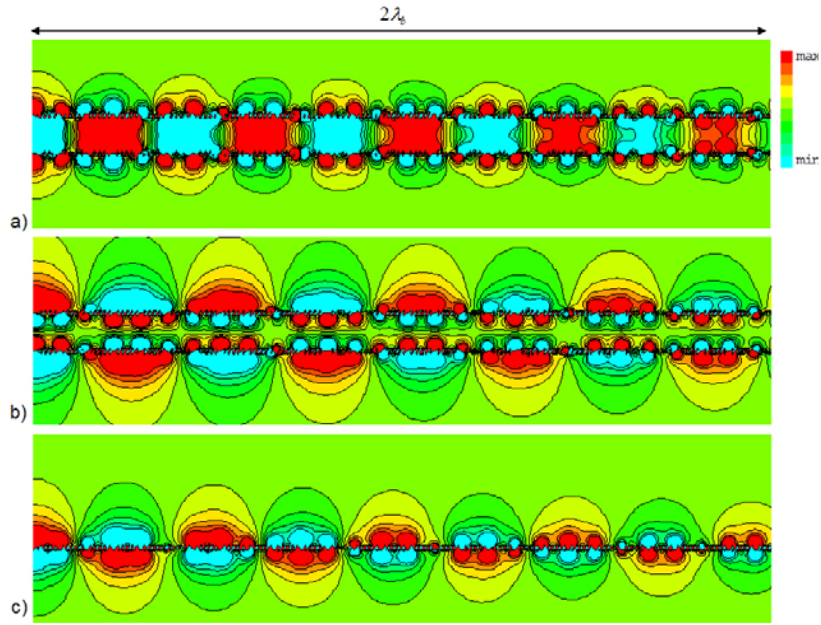

Figure 5 – (Color online). Similar to Fig. 4, but for frequency $f = 680\,THz$.

From the previous examples, it is evident that in this regime the modes guided by the parallel chains are quasi-longitudinal with a spurious transverse polarization, arising from the coupling, which is nearly 90° out of phase with respect to the longitudinal polarization. In Fig. 6 , for the parallel chains of Fig. 2 and 3 we have calculated the level of transverse cross-polarization induced on the particles due to coupling, as a function of frequency. It is evident that its level increases for closer chains, as expected, and it is larger for antisymmetric modes. In the region of enhanced absorption that we have noticed in Fig. 3, the corresponding level of cross-polarization is also very high, at some frequencies even higher than the longitudinal polarization, noticeably affecting the chain guidance. The coupling is minimal near the light line and in the leaky-wave and stop-band regimes, whereas it hits its maximum somewhere inside the guidance region, whose position in frequency varies depending on the distance between the chains.

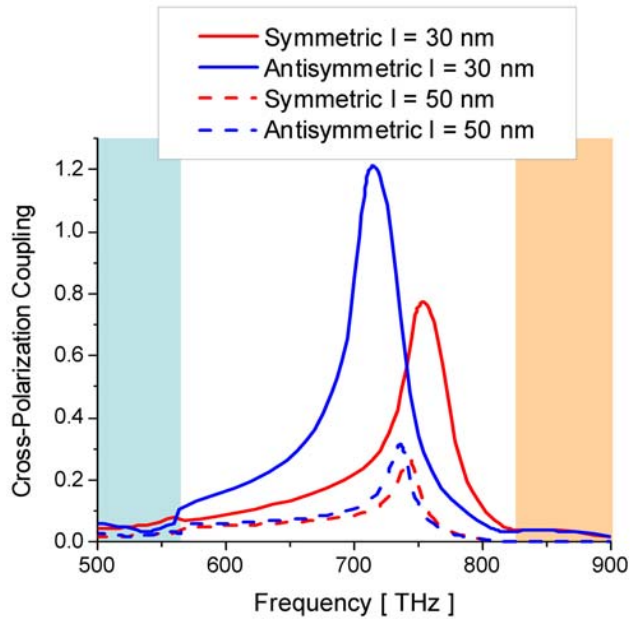

Figure 6 – (Color online). Magnitude of the transverse cross-polarization for the chains of Figs. 2 and 3, operating in their quasi-longitudinal forward-wave regime.

*b) Quasi-transverse y -polarized propagation (backward modes)*

We have reported in [10] that a single isolated linear chain may also support transversely polarized guided modes, satisfying the exact dispersion relation $T = 0$. In this case, the condition on the particle polarizability is:

$$\bar{d}^3 \bar{\alpha}_{min}^{-1} < \bar{d}^3 \operatorname{Re}\left[\bar{\alpha}^{-1}\right] < -3\left[Cl_3\left(\bar{d}+\pi\right)+\bar{d}\, Cl_2\left(\bar{d}+\pi\right)-\bar{d}^2 Cl_1\left(\bar{d}+\pi\right)\right], \qquad (10)$$

where $\bar{\alpha}_{min}^{-1}$ is defined in [10]. In this regime the chain always supports two modes, both with the same transverse polarization: one is guided along the chain and has backward-wave properties, the other is weakly guided, with forward-wave properties and $\operatorname{Re}\left[\bar{\beta}\right] \simeq 1$ (this eigenmode is basically a simple plane wave traveling in the background region, weakly polarizing the nanoparticles. This is not of interest for guidance purposes [10], but it is still reported here for sake of completeness). For the geometry at hand, transversely-polarized propagation is supported over the frequencies between $650 THz$ and $800 THz$, in part overlapping with the longitudinally-polarized regime, as reported in Fig. 5 (thin black line), consistent with Eq. (10). Remarkable differences are noticed between longitudinal and transverse polarization: the confined transverse mode is backward in nature, explaining the negative slope of $\operatorname{Re}\left[\bar{\beta}\right]$ versus frequency and the negative sign of $\operatorname{Im}\left[\bar{\beta}\right]$. Correspondingly, the bandwidth is more limited and losses are higher, as is usually the case for backward-wave waveguides. As a consequence, the leaky-wave operation arises now at the upper boundary of the guided regime (right shadowed region in Fig. 5) [18], whereas the Bragg stop-band is positioned at the lower-end of the guidance band.

Due to the modal coupling in the $xy$ plane, when the coupling between parallel chains is considered the quasi-transverse modes still satisfy the dispersion relations (8) given in the previous section and the polarization eigenvectors obey the same relations (9). It should be noticed, however, that in this regime the modes are quasi-transverse, and therefore the antisymmetric mode now corresponds to parallel $y$-polarized chains, whereas the symmetric mode supports anti-parallel polarization along $y$, consistent with (9).

Figure 7 reports the dispersion of symmetric and antisymmetric quasi-transverse modes for $l = 50\,nm$. Once again, the relatively small coupling between the chains produces a minor perturbation of the original backward-wave purely-transverse mode, which causes the antisymmetric mode to have slightly lower real and slightly larger imaginary part of $\bar{\beta}$. Conversely, the symmetric mode supports slightly larger values of $\text{Re}\left[\bar{\beta}\right]$. As in the previous section, the coupling is stronger near the light line and in the leaky-wave region, as expected. Also in this scenario the symmetric operation allows longer propagation lengths, even if in this case the $y$-polarized currents are oppositely oriented. Compared to quasi-longitudinal forward modes, the propagation length is sensibly reduced. Of course, following the results in [10], the propagation length may be somewhat increased and optimized by increasing the size of the nanoparticles and/or reducing the interparticle distance $d$. These results show that the pairing between two parallel chains may substantially increase the propagation length of these backward-wave optical nanowaveguides, which may be of interest for several applications.

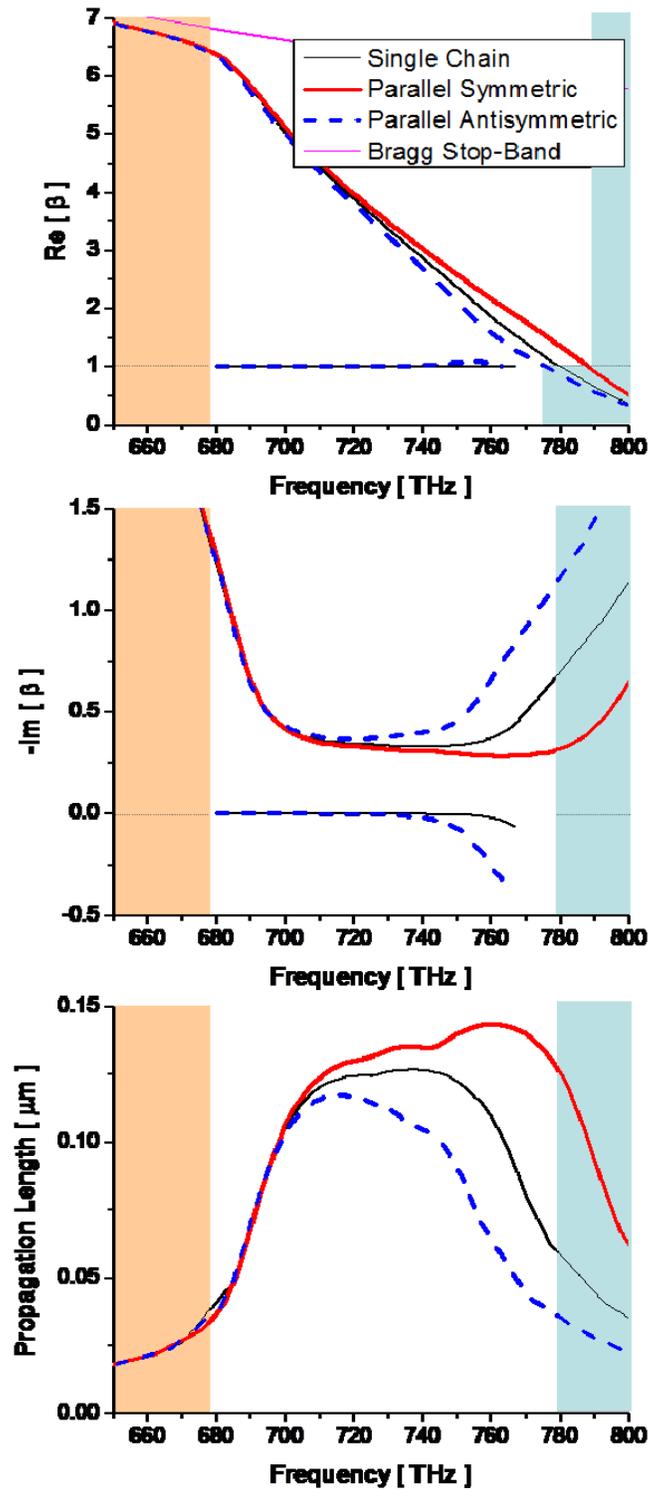

Figure 7 – (Color online). Similar to Fig. 2, but here in the quasi-transverse $y$ – polarized regime. Here the interchain distance is $l = 50\,nm$.

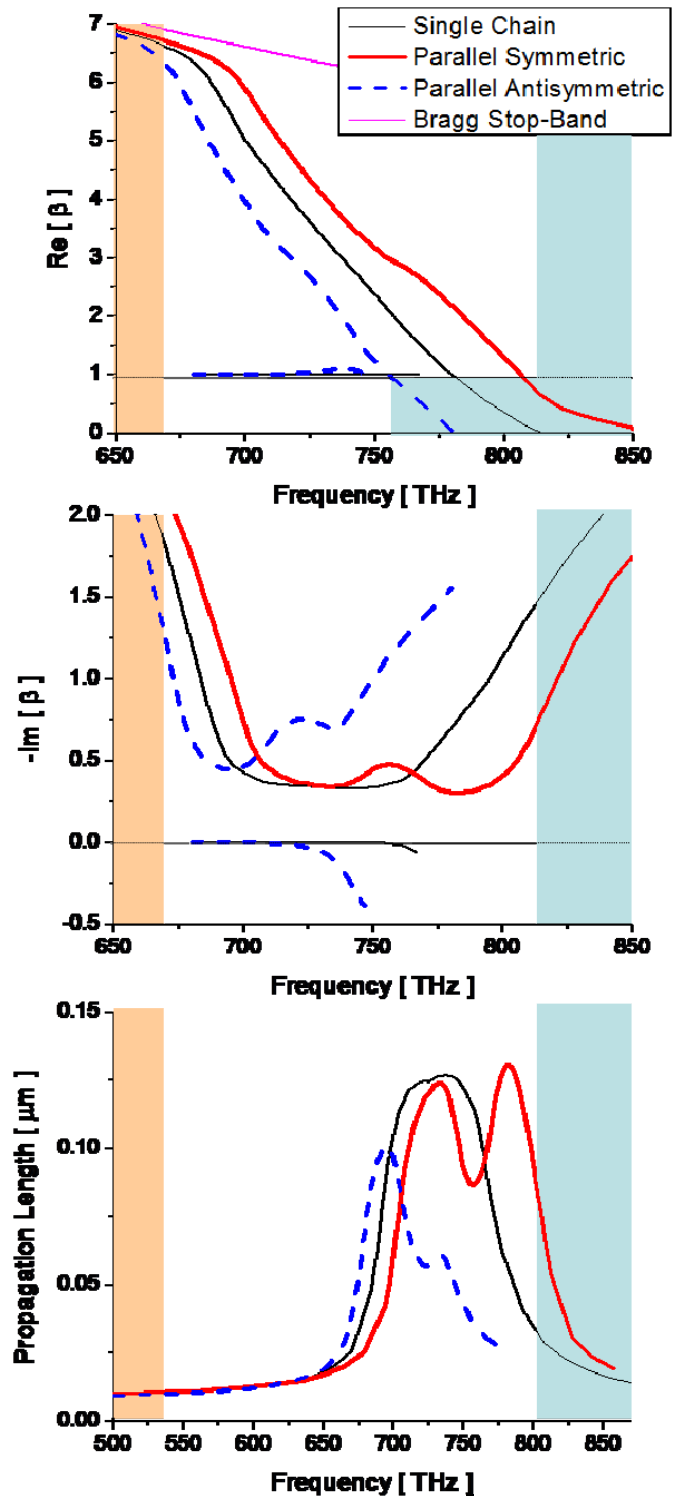

Figure 8 – (Color online). Similar to Fig. 7, but for $l = 30\,nm$.

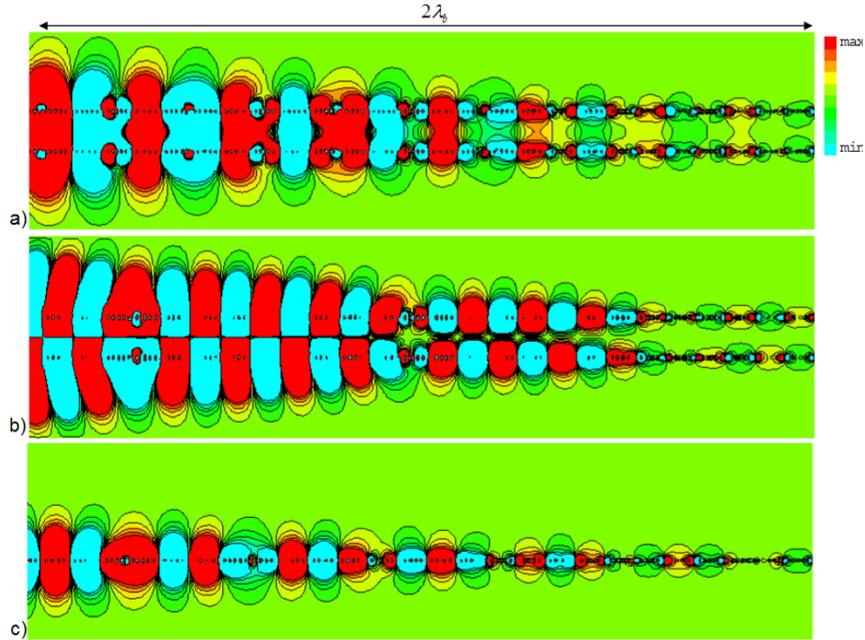

Figure 9 – (Color online). Magnetic field distribution (snapshot in time) for the chains of Fig. 8 at frequency $f = 700\,THz$. (a) antisymmetric mode, (b) symmetric mode, (c) isolated chain.

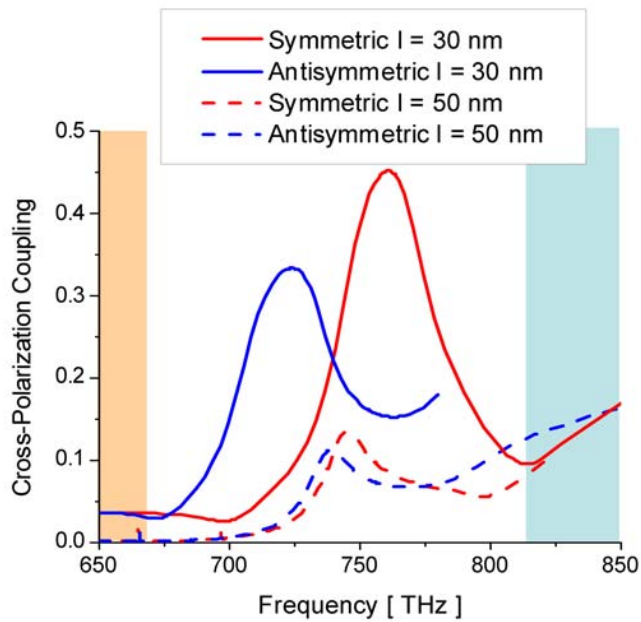

Figure 10 – (Color online). Amplitude of the longitudinal cross-polarization for the chains of Figs. 7 and 8, operating in the quasi-transverse regime.

Figure 8 reports analogous results in the case of closer chains ($l = 30\,nm$). Also in this case, the perturbation from the isolated chain is stronger and the bandwidth of backward operation may be substantially increased by using two parallel chains in the symmetric mode. Here leaky modes are proper in nature and therefore Eqs. (3) also apply to this regime in the way they are written. Both in Fig. 7 and 8, for completeness, we have also reported the modal branch associated with the weakly guided forward-wave transverse mode, which is located very close to the light line. Consistent with its forward-wave properties, $\text{Im}[\bar{\beta}] > 0$ for this mode. As outlined above, this mode is of minor interest for guidance purposes, since it is a minor perturbation of a plane wave traveling in the background region, very weakly affected by the presence of the chains. It is noticed, as expected, that this second branch is present only for the antisymmetric modes, whose $y-$polarization is in the same direction for both chains.

Figure 9 shows the magnetic field for these backward-wave modes as in Fig. 8 at the frequency $f = 700 THz$. In the antisymmetric case (Fig. 9a) $\mathbf{p}_1 = (0.076 - 0.12i)\hat{\mathbf{x}} + \hat{\mathbf{y}}$, $\mathbf{p}_2 = -(0.076 - 0.12i)\hat{\mathbf{x}} + \hat{\mathbf{y}}$ and $\bar{\beta}_{asym} = 3.95 - i0.46$; in the symmetric case $\mathbf{p}_1 = (0.02 - 0.016i)\hat{\mathbf{x}} - \hat{\mathbf{y}}$, $\mathbf{p}_2 = (0.02 - 0.016i)\hat{\mathbf{x}} + \hat{\mathbf{y}}$ and $\bar{\beta}_{sym} = 5.88 - i0.72$; for the isolated chain $\bar{\beta}_{single} = 5.036 - i0.42$. The field distributions in some senses resemble the one for quasi-longitudinal modes, but the presence of a dominant transverse polarization does not allow an analogous strong transmission-line confinement in this backward-wave regime for the antisymmetric modes. Still, the plots confirm that relatively long backward-wave propagation (over one wavelength) is achievable using coupled parallel chains.

Figure 10 reports the level of longitudinal cross-polarization for the chains of Figs. 7 and 8. In this scenario, the cross-polarization is in general lower than for quasi-longitudinal modes and it is stronger for symmetric modes. Once again, the cross-polarization is stronger for closely coupled chains and it has some resonant peaks in the middle of the guidance region, for which the damping is increased correspondingly.

*c) Purely transverse z-polarized propagation (backward modes)*

When the chains are polarized along $\hat{\mathbf{z}}$ the supported modes are purely transverse, consistent with (4). Due to the symmetry, the properties for isolated chains are identical to those described in the previous section, and therefore here we discuss how the coupling may affect differently the backward-wave guidance properties in this polarization. The coupling coefficient $C_{zz}$ splits the transverse modal branch of propagation into two modes, with dispersion relations:

$$\begin{aligned} sym: & \quad T + C_{zz} = 0 \\ antisym: & \quad T - C_{zz} = 0 \end{aligned} \qquad (11)$$

providing the following constraints on the polarization eigenvectors for the two chains:

$$\begin{aligned} sym: & \quad \mathbf{p}_1 \cdot \hat{\mathbf{z}} = \mathbf{p}_2 \cdot \hat{\mathbf{z}} \\ antisym: & \quad \mathbf{p}_1 \cdot \hat{\mathbf{z}} = -\mathbf{p}_2 \cdot \hat{\mathbf{z}} \end{aligned} \qquad (12)$$

Figure 11 reports the dispersion of symmetric and antisymmetric transverse modes for $l = 50\,nm$. Here, for the same distance as in Figs. 2 and 7, the coupling perturbs the propagation properties even less as compared to the isolated chains. Also in this case, symmetric modes allow slightly longer propagation lengths near the light line, where the coupling is stronger. Increasing the coupling ($l = 30\,nm$), as in Fig. 12, the perturbation is stronger, even if the trend is similar as in the previous scenario.

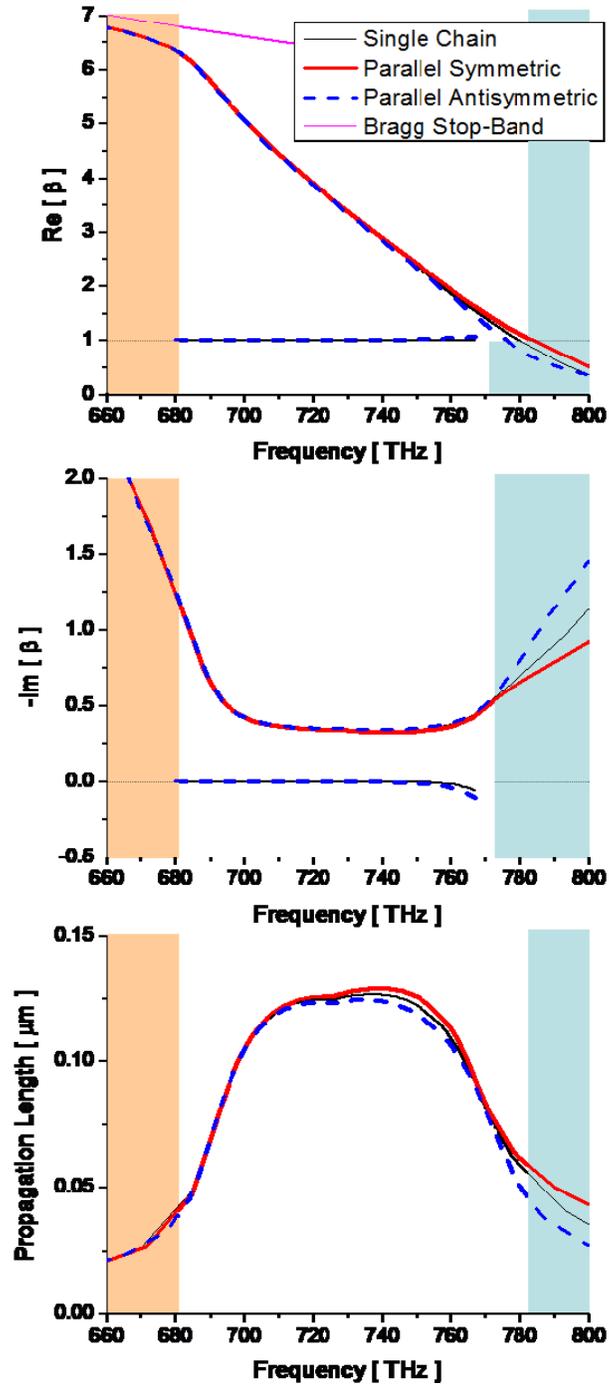

Figure 11 – (Color online). Similar to Fig. 2 and Fig. 7, but for transverse $z$ – polarized modes. Here the interchain distance is $l = 50\,nm$.

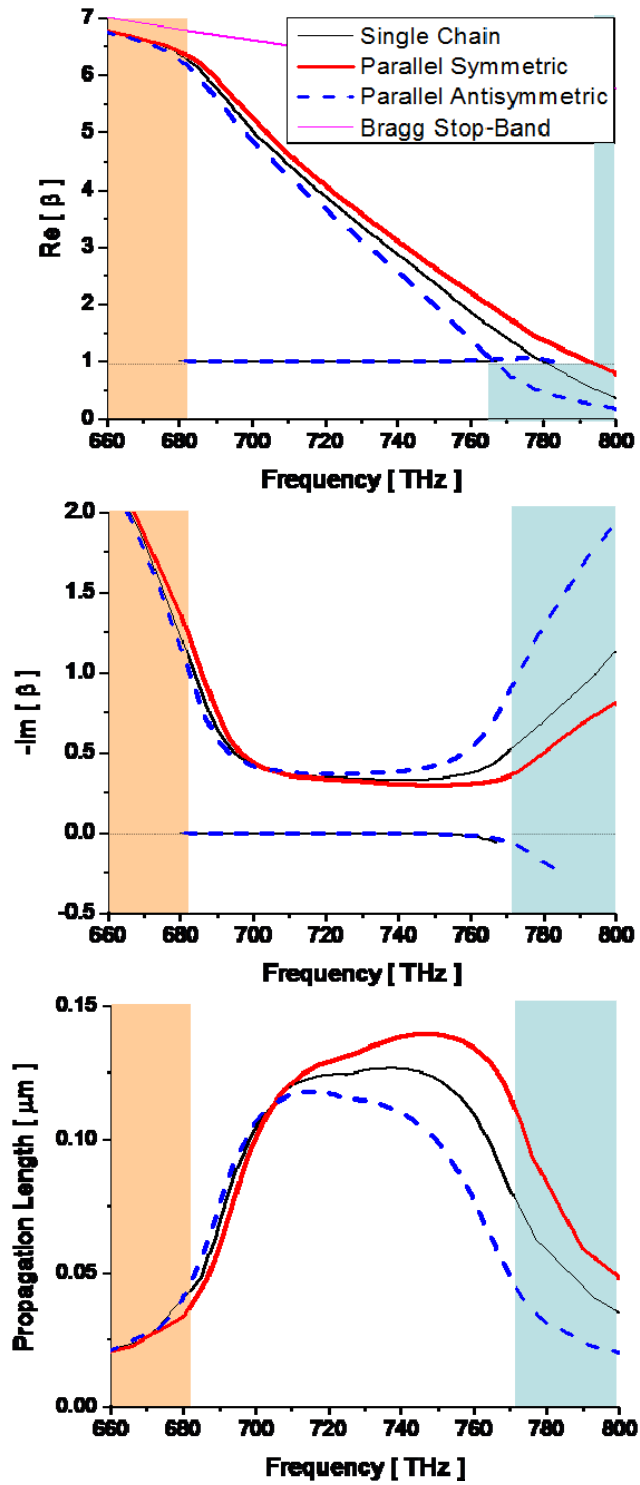

Figure 12 – (Color online). Similar to Fig. 11, but for $l = 30\,nm$.

Figure 13 reports the calculated orthogonal electric field distribution in the $xy$ plane for the modes of Fig. 11 at the frequency $f = 750 THz$. In this case the modes are purely transversely polarized and the guided wave numbers are respectively $\bar{\beta}_{asym} = 1.987 - 0.42i$, $\bar{\beta}_{sym} = 2.64 - 0.296i$, $\bar{\beta}_{single} = 2.36 - 0.34i$, consistent with Fig. 12. The field confinement in this polarization is not drastically different from that of an isolated chain, as evident from the figure, and the main advantage of using parallel chains may reside in the longer propagation distance of symmetric modes near the light line.

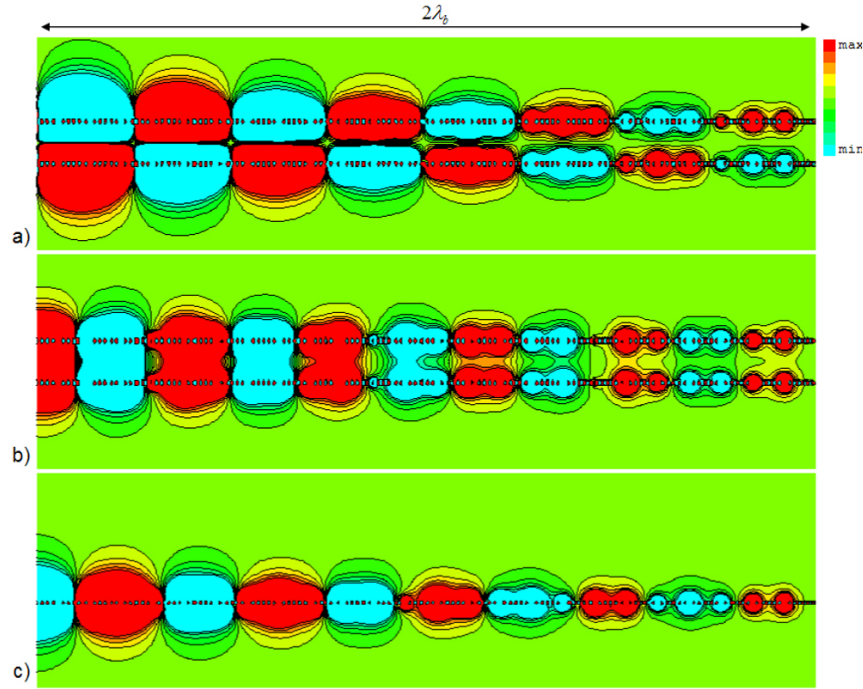

Figure 13 – (Color online). Electric field distribution (snapshot in time) for the chains of Fig. 12 at frequency $f = 750 THz$. (a) antisymmetric mode, (b) symmetric mode, (c) isolated chain.

## 4. Conclusions

We have presented here a fully general and complete theoretical formulation for the analysis of the dynamic coupling between two parallel linear chains of plasmonic nanoparticles operating as optical waveguides. These chains may support up to eight different guided modes with different polarization properties in the same range of frequencies, which we have fully analyzed here. We have shown that, compared to linear arrays, these waveguides may support longer propagation lengths and ultra-confined beams, operating analogously to transmission-line segments at lower frequencies. In particular, our results confirm that by operating near the light line with antisymmetric quasi-longitudinal modes we may achieve relatively long propagation lengths (of several wavelengths) and ultraconfined beam traveling, similar to a transmission-line, in the background region sandwiched between the two antisymmetric current flows guided by the chains. Our analysis has fully taken into account the whole dynamic interaction among the infinite number of nanoparticles, also considering presence of material and radiation losses and the frequency dispersion of the involved plasmonic materials.

*Acknowledgements*

This work is supported in part by the U.S. Air Force Office of Scientific Research (AFOSR) grant number FA9550-08-1-0220 to N. Engheta and by the U.S. Air Force Research Laboratory (AFRL) grant number FA8718-09-C-0061 to A. Alù.